\documentclass[%
 reprint,
 amsmath,amssymb,
 aps,
]{revtex4-2}

\usepackage{mathtools}
\usepackage[utf8]{inputenc}
\usepackage{float}
\usepackage{multirow}
 \usepackage{pifont}
\usepackage{xcolor}
\usepackage{hyperref} %Interactive references
\usepackage[all]{hypcap} %Shows exactly where the figure/table is located when clicking.
\hypersetup{colorlinks=true,linkcolor=blue,citecolor=blue} %Colourful interactive references.
\usepackage{graphicx}
\usepackage{enumitem}
\usepackage{afterpage}

\graphicspath{{figures/}}

\usepackage{nomencl}
\makenomenclature

\usepackage{etoolbox}
\usepackage{soul}
\usepackage{amsmath}
\renewcommand\nomgroup[1]{%
  \item[\bfseries
  \ifstrequal{#1}{A}{Acronyms}{%
  \ifstrequal{#1}{B}{Symbols}}%%
]}

\begin{document}
\preprint{APS/123-QED}

\pagenumbering{roman}

\title{The evolution of invasion patterns due to surfactant adsorption in anomalous pore distribution: Role of Mass Transfer and Laplace Pressure}% Force line breaks with \\

\author{Debanik Bhattacharjee}
\author{Guy Z. Ramon}
\author{Yaniv Edery}
\email{Corresponding author: yanivedery@technion.ac.il}
 
\affiliation{Department of Civil \& Environmental Engineering Technion - Israel Institute of Technology Haifa 32000 Israel}%

\begin{abstract}
\pagenumbering{arabic}

Immiscible two-phase flow in porous media occurs in many processes, such as enhanced oil recovery (EOR) as well as oil-spill and soil remediation. 
These processes involve a fluid displacing another immiscible fluid within the confines of a heterogeneous porous structure. 
The invasion pattern generally remains the same under constant conditions but can also evolve over time in the presence of surfactants, which alter the interfacial tension (IFT) and surface wettability. 
The dynamics under such conditions extend beyond the usual way in which such immiscible displacement is modeled.
Here, we develop a time-dependent pore network model (PNM) to simulate the effects of surfactant-induced IFT reduction on immiscible displacement driven by constant inlet pressure, with pressure drops across the network calculated using a random resistor network and mass conservation equations. 
Node-specific flux and velocity are derived using the Hagen-Poiseuille equation, and surfactant adsorption is modeled using the Langmuir isotherm, capturing its impact on fluid-fluid and solid-fluid interfaces within the invaded path.
 
Since the evolution of the invasion pattern comprises the cooperative mechanisms of surfactant mass transfer to the interfaces and the resulting changes in capillary and Laplace pressures, we employ two strategies to quantify this complex feedback behavior: mass transfer-based, introducing a mass transfer timescale, and Laplace pressure-based, scaling with the inlet pressure. 
Results reveal that an anomalous or heavy-tailed pore throat distribution accelerates the onset of secondary invasions, which enhances the dominance of Laplace pressure. 
As the distribution becomes less anomalous or more symmetric, mass transfer becomes the dominant mechanism. 
This interplay highlights the intricate balance between mass transfer and capillary effects in governing the spatio-temporal evolution of immiscible fluid invasion. 

%By capturing these dynamics, our framework provides deeper insights into the complex mechanisms driving fluid displacement in heterogeneous porous media.

\end{abstract}

\maketitle
\section{Introduction}

The displacement of a defending fluid by an immiscible fluid in heterogeneous porous media is found in soil remediation \cite{Haigh1996,Urum2004,Silva2021,Conte2005,YanZHeng2007,Stelmack1999,Christofi2002,Mao2015,Fenibo2019,Parthipan2021}, oil spill remediation \cite{Prendergast2014,Joye2020,Board2019,Silva2022,Zhu2022,Shah2022,Adofo2022,Nikolova2021,Head2006,Ng2022} and Enhanced Oil Recovery (EOR) \cite{Zhu2018,ElHoshoudy2017,Haruna2020,AfzaliTabar2020,You2018,Sheng2015,Massarweh2020,Howe2015,Jing2021,Zhao2022}. 
In these processes, a resident fluid is displaced by an immiscible invading fluid, within the connected pore structure. 
This displacement requires that the pressure within the invading phase is higher than the local capillary pressure, eventually forming a connected path that allows for a continuous flow through the porous domain. 
As such, invasion pattern advancement is mainly dependent on the local pore structure and wettability in a framework known as invasion-percolation \cite{Wilkinson1983,Lenormand1983}. 
However, in heterogeneous porous media, the presence of anomalous transport, caused by broad, heavy-tailed distributions of pore throats and irregular connectivity, can significantly alter the invasion dynamics \cite{eliyahu-yakir_mixing_2024}, and any associated processes, such as chemical reactions \cite{biran_experimental_2024, shavelzon_shannon_2023}. 
Such anomalous transport can lead to preferential flow pathways, delayed breakthrough, and enhanced trapping of the defending fluid, making it crucial to consider in modeling immiscible displacement processes \cite{Dagan2022,edery_effect_2021,Berkowitz2000,Zhang2024,Edery2014,Rajyaguru2024}.

The local consideration of capillarity forms the basis of the invasion-percolation framework, aligning with the Pore Network Model (PNM) initially employed by Fatt in the 1950s to simulate fluid flow in porous media \cite{Fatt1956}. 
As PNM offers the distinct advantage of exact solution of multiple phases in specific pores while considering their wettability and size, it can implement the invasion percolation theory and simulate the invasion of the immiscible fluids in porous media.  This entirely depends on overcoming the local capillary pressure, which is a function of interfacial tension, wettability, and geometry. 
Applying this local criterion at the network scale leads to the invasion percolation model (IPM), which provides a simple algorithm for simulating the slow displacement of a wetting fluid by a non-wetting fluid in a porous medium.  
Within the IPM, a network consists of a distribution of pores and, as a result, a distribution of interfaces, which implies that a wide range of capillary pressures are present. 
The criteria for one fluid to ``invade" the pore space occupied by another fluid is that the threshold capillary pressure at each interface is exceeded by the inlet hydrostatic pressure, $\mathrm{p_{in}}$. 
This algorithm ensures that there is a connected path between all the interfaces, from a fixed inlet to the outlet, fulfilling the invasion criterion: $\mathrm{p_{i} < p_{in}}$. 
This allows the pore-scale heterogeneity, which is a common occurrence in soil, rocks and sediments \cite{Ye2015,Knackstedt2001,Trevisan2015}, known to have an effect on the invasion processes, to be modeled as a mesh of tubes with the same statistical size distribution as the sample, which follow an anomalous or heavy-tailed distribution in this study.

In this model, once the constant hydrostatic inlet pressure forms the invasion pattern, it does not evolve further, while the applied pressure becomes a pressure drop that maintains the invading fluid flow. 
However, soil and rocks are constantly subjected to contamination and exhibit chemical impurities that may alter both the solid surface wettability and the fluid-fluid interfacial tension. 
An important example of such a process is the presence of surfactants, which are often added into effluents to solubilize contaminants from the soil \cite{Haigh1996,Urum2004,Silva2021,Conte2005}. 
Over the past few years, a number of studies have looked into the effect of surfactants on immiscible displacement in porous media by altering interfacial tension and wettability through flooding experiments \cite{Mejia2019,Yang2021,She2021,Ramezanzadeh2022} or dynamic pore network modelling \cite{Qin2020,Hammond2012}. 
In the latter studies, the authors reported the mechanisms of surfactant adsorption as well as the importance of pore-scale heterogeneity on displacement mechanisms. 
However, to the best of our knowledge, there has been no numerical study which investigated how surfactant adsorption at interfaces, after the IPM phase, affects the spatio-temporal evolution of this initial invasion, while also systematically considering the role of anomalous heterogeneity of the porous media, particularly across a statistically significant number of random network configurations. 

The focus of the present work is to investigate the role of surfactants on the invasion dynamics after the IPM framework established the initial invasion. This invasion dynamics is numerically investigated using a PNM algorithm that considers surfactant transport, adsorption, and subsequent changes to interfacial tension. 
Since surfactants reduce the surface tension, they reduce the Laplace pressure difference across the capillary interface, which in turn increases the extent of invasion and alters the flow by updating the local pressure drop (while the global pressure difference remains constant). 
To capture the influence of surfactants on secondary invasion, we extend the IPM by calculating viscous pressure drops and fluxes in the PNM. The local surfactant concentration is then established using a  Surfactant Transport Model (STM), followed by a Surfactant Adsorption Model (SAM) that modifies interfacial tension and contact angle based on surfactant diffusion and Langmuir adsorption kinetics.

As surfactants migrate from the bulk to interfaces, capillary pressures decrease, facilitating further invasion and evolving flow patterns. 
This dynamic redistribution of surfactants leads to a cooperative interaction between mass transfer and changing Laplace pressures. Inspired by soil water retention models \cite{Kosugi1994,Kosugi1996}, we develop a framework to systematically analyze invasion dynamics through system parameters linked to mass transfer and dynamic Laplace pressure, highlighting the complementary insights each approach provides.

\section{Methodology}

\subsection{Pore Network Model (PNM)} 

\subsubsection{Geometry} \label{geometry}

\begin{figure*}
\centering
\includegraphics[width=0.8\linewidth]{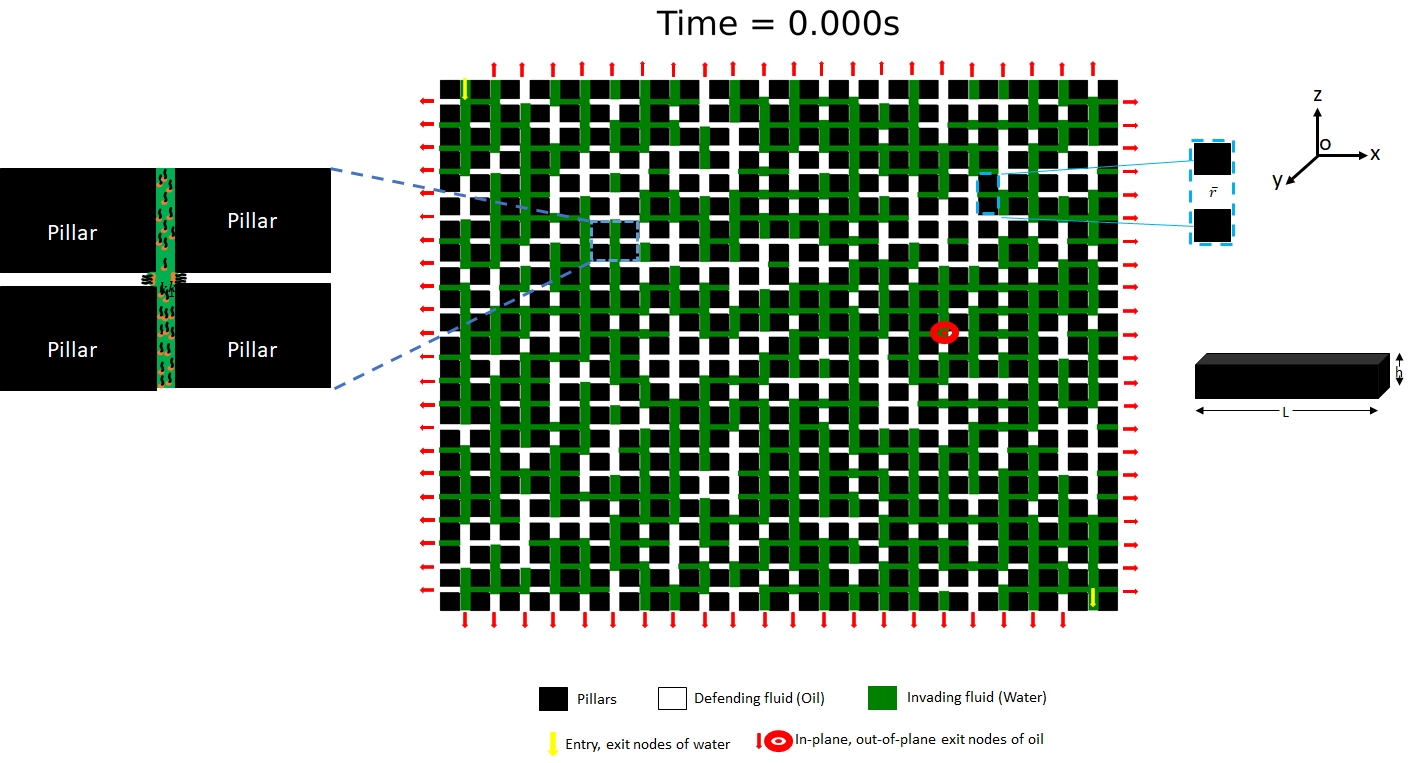}
\caption{A typical layout of a 23x23 pillar array depicting primary invasion pattern.}
\label{fig:1} 
\end{figure*}

An efficient way to implement the IPM through the PNM is by considering a pillar array (represented by black squares) that forms the pore network (Fig.~\ref{fig:1}), where the junction between the pillars can be classified as pore bodies, and the pillar edges classified as pore throats. 
These pore throats follow a Gamma distribution function, which has been used in previous studies \cite{Ma2024,Wang2020}, and their connectivity is established using an adjacency matrix \cite{Gostick2016}. 
For the present calculations, the mean pore throat size is set as $50\ \mu\mathrm{m}$, with the variance ranging between $24-1251 \ \mu \mathrm{m^2}.$ 
Further details on the adjacency matrix and network numbering scheme can be found in the Supplementary Information. 

We consider a defending fluid (oil) and an invading fluid (water), the properties of which are provided in Table.~\ref{table:1}. 
To reduce computational time, we choose a fixed inlet and an outlet, both of which are marked with yellow arrows (Fig.~\ref{fig:1}). 
The inlet pressure has a finite, non-zero value and the outlet pressure is assumed to be atmospheric. 
The invasion pattern is obtained from the adjacency matrix.  

\subsubsection{Quasi-static Invasion Percolation}
 
We use a quasi-static invasion percolation algorithm to study how an immiscible fluid phase replaces a defending fluid within the pore network geometry \cite{JoekarNiasar2013,JoekarNiasar2008,JoekarNiasar2010}. 
The invasion percolation algorithm is implemented on a random 2D pore network, as shown in Fig.~\ref{fig:1}. 
Following the distribution of pore throats depicted in \ref{geometry}, an inlet pressure is chosen to be the capillary pressure corresponding to the largest pore throat at the interface of two phases. 
The capillary pressure for a rectangular cross-section is given by \cite{Juncker2002}
 
%We consider a rectangular cross-section of each pillar. 
%Thus, the flow in the network is representative of a flow in a rectangular microchannel. 
%The capillary pressure and hydraulic resistance with a rectangular cross-section are given, respectively, by \cite{Juncker2002}, shown by the following equation. 

\begin{equation} 
\mathrm{p}_{\mathrm{i}} = -{\mathrm{\sigma}}\Bigg(\frac{\cos\alpha_{\mathrm{b}} + \cos\alpha_{\mathrm{t}}}{\mathrm{h}} + \frac{\cos\alpha_{\mathrm{l}} + \cos\alpha_{\mathrm{r}}}{r_{\mathrm{i}}}\Bigg),
\label{eq:1}
\end{equation}

where $\mathrm{p_i}$ is the capillary pressure of node `i', $\sigma$ is the interfacial tension, $\alpha_\mathrm{b0},\alpha_\mathrm{t0},\alpha_\mathrm{l0},\alpha_\mathrm{r0}$ are contact angles at the bottom, top, left and right of the rectangular pillar, respectively, $h$ is the height of the microchannel, and $\mathrm{r_i}$ is the pore throat size. 
The algorithm searches the interfaces marked by the adjacency matrix that possess values smaller than or equal to the inlet pressure, and advance the invasion utill it reaches the fixed outlet of the invading fluid, thus ensuring a minimal connected path from inlet to outlet. 

\subsubsection{Hydrodynamic flow}

Once a connected path is formed, the applied static pressure at the inlet establishes a pressure drop that drives the fluid flow in the connected path, with individual hydraulic resistances given by  

\begin{equation} 
\mathrm{R_i^H} = \frac{12 \, \mu \mathrm{L}}{\mathrm{h^3 r_i} \left(1 - 0.63 \left(\mathrm{\frac{h}{r_i}}\right)\right)},
\label{eq:2}
\end{equation}

where $\mathrm{R_{i}^H}$ is the hydraulic resistance of node `i', h is the height of the microchannel (while 12 stems from the upper and lower plate drag force, see \cite{homsy1987viscous}), $\mathrm{r_i}$ is the pore throat size, $\mu$ is the dynamic viscosity of the fluid, and L is the length of each pillar. 

%To initiate the simulation, an inlet pressure is required which is chosen to be the capillary pressure corresponding to the largest pore throat. The algorithm searches the interfaces marked by the adjacency matrix with values less than or equal to inlet pressure and progresses till it reaches the fixed outlet of the invading fluid, thus ensuring a minimal connected path from inlet to outlet. Once a connected path is formed, the applied static pressure at the inlet now forms the inlet pressure, thus establishing a pressure drop due to the fluid flow in the connected path. Once the connected path from inlet to outlet is formed, the final inlet pressure, which quasi-statically advanced the invasion, establishes the flow through local viscous pressure drops at each pore throat. 

We calculate the local fluid flux and viscous pressure by writing a mass balance equation (eq.~\ref{eq:3}) analogous to Kirchoff's law, which states that the amount of flux entering and leaving the node must be equal. 
This provides the pressure drop values, $\mathrm{\delta P_{ij}}$ across each node. 
The Hagen-Poiseuille equation is then used to relate the flux, Q, to $\mathrm{\delta P_{ij}}$ (eq.~\ref{eq:4}) using hydraulic resistance, $\mathrm{R_{ij}^H}$ defined in eq.~\ref{eq:2}. 
Eventually, the velocity, $\mathrm{V_{ij}}$ is calculated as follows.

\begin{equation} 
\mathrm{\sum_{j}\frac{P_i}{R_{i}^H}}=0
\label{eq:3}
\end{equation}
\begin{equation} 
\mathrm{Q_{ij}} = \mathrm{\frac{\delta P_{ij}}{R_{ij}^{H}}}
\label{eq:4}
\end{equation}
\begin{equation} 
\mathrm{V_{ij}} = \mathrm{\frac{Q_{ij}}{hr_i}}
\label{eq:5}
\end{equation}
where $\mathrm{Q_{ij}}$ is the flux,  $\mathrm{\delta P_{ij}}$ is the viscous pressure drop, $\mathrm{V_{ij}}$ is the velocity of the invading fluid connecting nodes `i',`j'.

\begin{table*}
\centering
\includegraphics[width=0.8\linewidth]{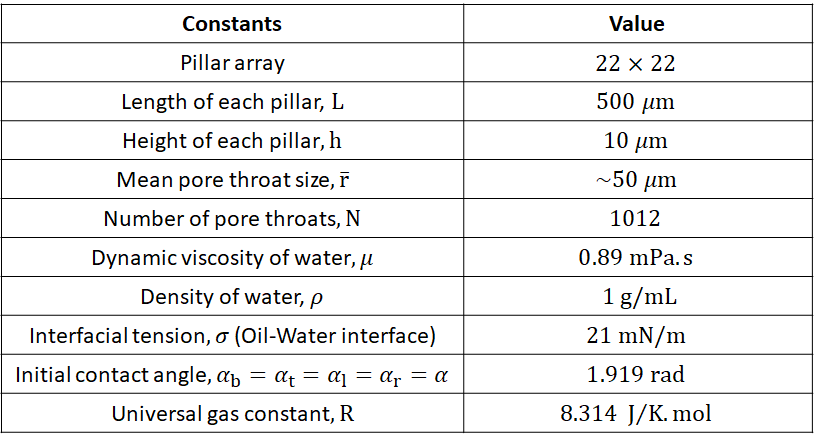}
\caption{Various parameter values used in the simulation.}
\label{table:1} 
\end{table*}

\subsection{Surfactant Transport Model (STM)}

Once the IP path for the invading phase is formed, we introduce surfactants to the invading fluid. 
Since there is a flow path between inlet and outlet, surfactants in the feed flow migrate to the oil-water interfaces. 
We assume that surfactant reaches the interfaces by a combination of convection and diffusion, simplified by the introduction of a mass transfer coefficient. 
At a location close to the interface, we have 
\begin{equation} 
\mathrm{\frac{dC_I}{dt}=\frac{k}{L_{char}}({C_B-C_I})},
\label{eq:6}
\end{equation}

where $\mathrm{C_B}$ is the bulk concentration of the surfactant, $\mathrm{C_I}$ is the concentration close to the interface, `I', $\mathrm{L_{char}(=(\frac{V}{S}))}$ is the characteristic channel length scale defined as inverse of the ratio of local surface area, S, to volume, V, and k is the mass transfer coefficient. 
Integrating eq.~\ref{eq:6}, subject to the initial condition $\mathrm{C_I} = 0$ at $\mathrm{t=0}$, yields 

\begin{equation} 
\mathrm{{C_I}=C_B(1-exp(-\frac{k}{L_{char}}t))}.
\label{eq:7}
\end{equation}

To estimate the local mass transfer coefficient corresponding to an interface `I', we use the following mass transfer correlation,
\begin{equation} 
\mathrm{{Sh_I}={Pe_{I}^{1/3}}=\frac{k_{I}}{D/d_{I}}},
\label{eq:8}
\end{equation}

where $\mathrm{Sh_I,Pe_I,D,d_I}$ are the local Sherwood number, Peclet number, the diffusion coefficient, and the local hydraulic length of the geometry under consideration, respectively. 
Another observation is that the system has two interfaces: solid-fluid ($\mathrm{ISF}$) and fluid-fluid ($\mathrm{IFF}$). 
As a result, $\mathrm{C_I}$ can be rewritten as 

\begin{equation} 
\mathrm{C_{I}=n_{ISF}C_{ISF}+n_{IFF}C_{IFF}, }
\quad \mathrm{n_{ISF}+n_{IFF}=1.0}
\label{eq:9}
\end{equation}

where $\mathrm{n_{ISF},n_{IFF}}$ represent the fractions of concentration close to the solid-fluid (SF) and fluid-fluid (FF) interface, respectively, and, $\mathrm{C_{ISF}, C_{IFF}}$ are the concentration close to the solid-fluid and fluid-fluid interface, respectively. 
We assume that $\mathrm{n_{ISF}}$ is the ratio of the solid-fluid interface length to the total interface length, including solid-fluid and fluid-fluid.

\subsection{Surfactant Adsorption Model (SAM)}

While the STM establishes the surfactant concentration per location, the surfactant adsorption, which occurs at the fluid-fluid interfaces, follows the chemical equilibrium and alters both the interfacial tension and the wettability, i.e., contact angle, as described hereafter.

\subsubsection{Interfacial tension based on concentration close to the fluid-fluid interface}

To account for IFT changes at the fluid-fluid interfaces, we consider the Langmuir isotherm, for which, at constant temperature, single site occupancy can be expressed as

\begin{equation} 
\mathrm{\frac{\theta}{1-\theta}=K_LC_{IFF}}
\label{eq:10}
\end{equation}

where $\mathrm{\theta}$ is the fractional coverage of the oil-water interface by the surfactant molecules, and $\mathrm{K_L}$ is the Langmuir equilibrium constant. 
To relate the interfacial tension, $\sigma$ to $\mathrm{C_{IFF}}$, a common form of Gibbs' adsorption isotherm is considered.

\begin{equation} 
\mathrm{\frac{d\sigma}{dlnC_{IFF}}=-RT\Gamma=-RT\theta\Gamma_{max}}
\label{eq:11}
\end{equation}

where R is the universal gas constant, T is the temperature, $\Gamma$ is the adsorption at a certain coverage, $\Gamma_{max}$ is the maximum adsorption capacity.

Substituting $\theta$ from eq.~\ref{eq:10}, we get

\begin{equation} 
\mathrm{\frac{d\sigma}{dlnC_{IFF}}=-RT\Gamma_{max}\bigg(\frac{K_LC_{IFF}}{1+K_LC_{IFF}}\bigg)}.
\label{eq:12}
\end{equation}

Following \cite{Belton1976} and assuming $\mathrm{C_{IFF0}(t=0)=0}$, eq.~\ref{eq:12} is simplified into

\begin{equation} 
\mathrm{\sigma_1-\sigma_0=-RT\Gamma_{max}ln\bigg({1+K_LC_{IFF1}}\bigg)}
\label{eq:14}
\end{equation}

\subsubsection{Contact angle based on concentration close to the solid-fluid interface}

In our case, surfactants change the wettability by decreasing the contact angle as more and more surfactants are adsorbed at the solid-fluid interface, transforming the interface from hydrophobic to hydrophilic. For brevity, we assume that at any solid-fluid interface,

\begin{equation} 
\mathrm{\alpha_{itSF}=\alpha_{i0SF}-\frac{C_{ISF}*MW}{\rho}*100}
\label{eq:15}
\end{equation}

where $\mathrm{\alpha_{i0SF}}$ is the initial contact angle of the interface `i', $\mathrm{\alpha_{itSF}}$ is the contact angle at a later time t after surfactant adsorption at the solid-fluid interface, `MW' is the molecular weight of the surfactant, $\rho$ is the density of water.

\begin{table*}
\centering
\includegraphics[width=0.8\linewidth]{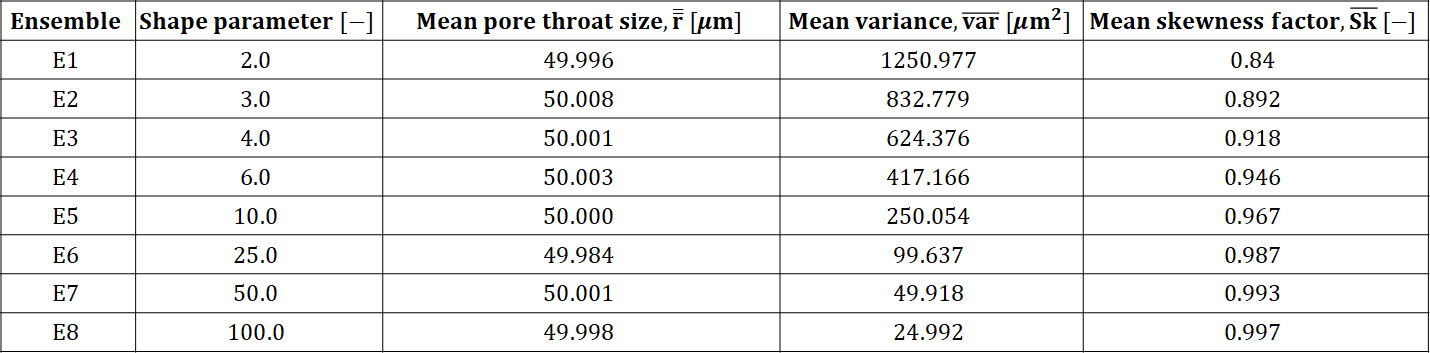}
\caption{Different ensembles used in the simulations.}
\label{table:2} 
\end{table*}

\subsection{Integration of Surfactant Transport and Adsorption Model into the Pore Network Model}

\begin{figure*}
%\begin{sidewaysfigure*}
\centering
\includegraphics[width=0.8\linewidth]{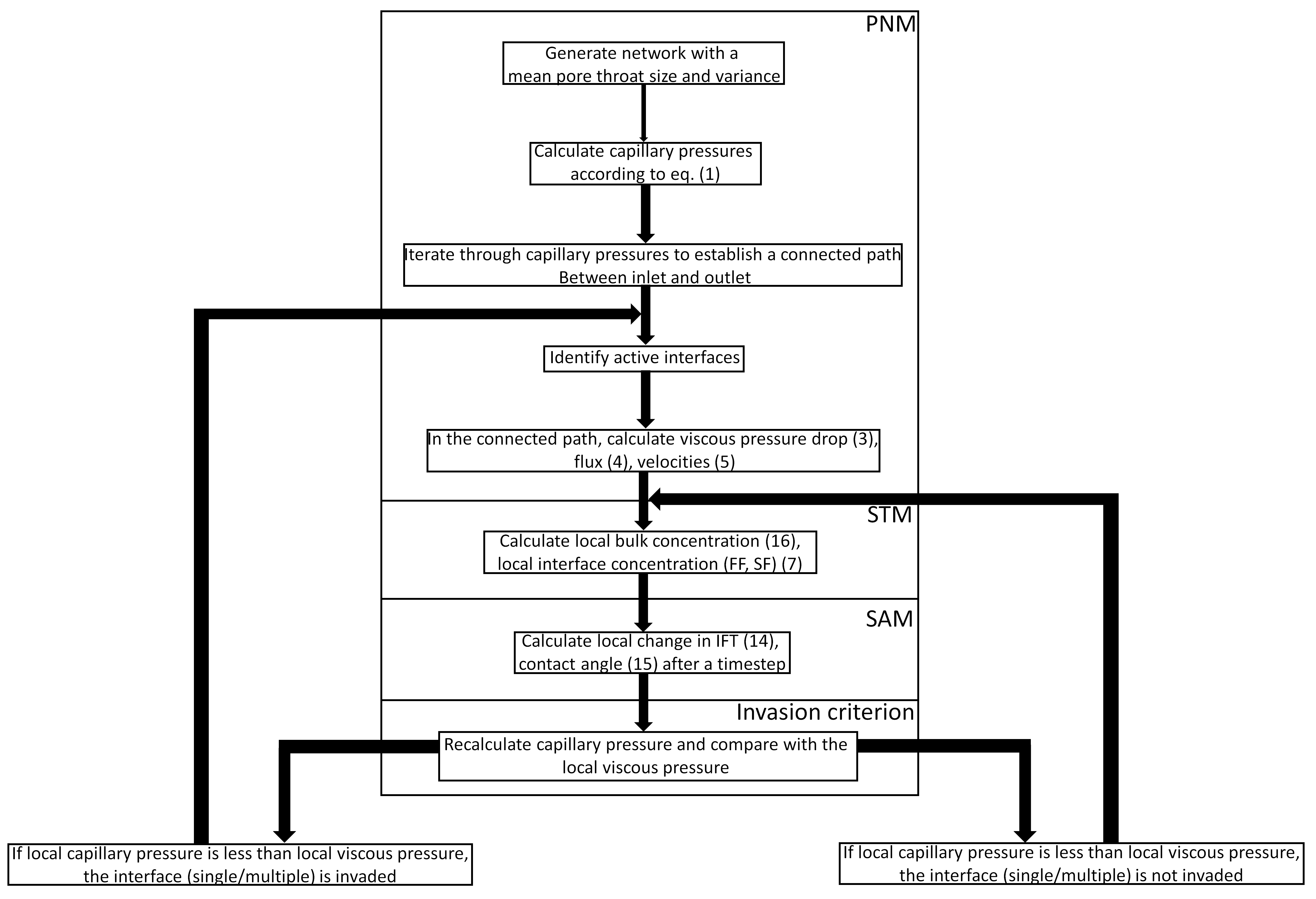}
\caption{A flowchart displaying how different models are integrated to study the surfactant driven invasion dynamics.}
\label{fig:2} 
%\end{sidewaysfigure*}
\end{figure*}

The important physical parameter calculated from the PNM is the flux, Q, through which we calculate the amount of bulk surfactant available in the vicinity of every fluid-fluid interface. Using STM, it is possible to calculate the surfactant concentration close to the fluid-fluid and solid-fluid interfaces. 
By subsequently employing the SAM for the updated surfactant concentration, the change in IFT and contact angle are found, which alter the capillary pressure. 

At the network level, we follow this algorithm to study the spatio-temporal surfactant-driven invasion dynamics, which we will refer to henceforth as "secondary invasion", for heterogeneous pore networks (also depicted in Fig.~\ref{fig:2}).
1. At time, t=0, we begin with an initial bulk concentration, $\mathrm{C_{B0} = 2} \, \text{mM}$. 
Applying a mass balance at every node within the invasion pattern, through which flow occurs, we have
\begin{equation} 
\mathrm{\sum_{j} C_{Bj}Q_{ij} - C_{Bi}Q_{ij}=0},
\label{eq:16}
\end{equation}

where $\mathrm{C_{Bj}}$ is the bulk concentration at node `j', $\mathrm{Q_{ij}}$ is the flux directed from node `i' to `j'. 
Thus, we have different bulk concentrations for every node.

2. Since we know $\mathrm{C_B}$, it is possible to find $\mathrm{C_I}$ based on eq.~\ref{eq:7} for every interface. 
Using eq.~\ref{eq:16}, we calculate the reduction in interfacial tension for the specific interface at each timestep. 
In a similar fashion, we obtain the change in contact angle, $\alpha$, for this interface.

3. At a given timestep, surfactant adsorption lowers the interfacial tension and, therefore, the capillary pressure. 
Furthermore, the contact angle is also reduced, owing to the adsorption of surfactants at the solid-fluid interface. 
This means that some interfaces will be invaded if the local viscous pressure exceeds the reduced local capillary pressure. 
As we have more invaded regions, the algorithm returns to step 1 to recalculate the bulk concentration at every node, and the cycle continues until no more interfaces are left to be invaded.
In this manner, there is a feedback loop between PNM, STM, and SAM, depending on the invasion criterion, which continues until a steady-state is reached, where there is a balance between the rates of fluid invasion and displacement, as shown in Fig.~\ref{fig:2}.

\subsection{Analysis} \label{analysis}

We analyse the spatio-temporal dynamics inspired by a framework developed to study soil water retention models, where for a Gaussian distribution of pore sizes, the relation between the applied pressure and the retention curve follows an error function \cite{Kosugi1994,Kosugi1996}. Our simulation tracks the interfaces as they are continuously invaded as a result of surfactant adsorption over time. 
This can be explained by scaling the time with a mean mass transfer timescale as shown in eq.\eqref{eq:17} which we term as Mass transfer (MT) analysis, governed by 

\begin{equation}
\mathrm{f_{t\chi} = \left( f_{\infty \chi} - f_{0\chi} \right) \phi \left( \frac{\frac{t_{\chi}}{\overline{\tau_{\chi}}} - \mu_{t\chi}}{\sigma_{t\chi}} \right) + f_{0\chi}}.
\label{eq:17}
\end{equation} 
This continuous invasion can also be seen as the change in the mean Laplace pressure in relation to the available interfaces, which is analysed using eq.\eqref{eq:18}, termed as the Laplace pressure (LP) based analysis, 

\begin{equation}
\mathrm{f_{p\chi} = \left( f_{\infty \chi} - f_{0\chi} \right) \phi \left( \frac{\frac{{\Delta}{\overline p_{\chi}}(t)}{P_{in,\chi}} - (-\mu_{p\chi})}{\sigma_{p\chi}} \right) + f_{0\chi}} %\chi=1,2,3,\ldots}
\label{eq:18}
\end{equation}

At a given time, the invaded fraction is represented by the number of interfaces invaded, divided by the total number of pore throats in the network (1012, in this case).

The quantities $\mathrm{f_{t\chi}},\mathrm{f_{p\chi}} [-]$ are the invaded fractions as a function of time, $\mathrm{t[s]}$, based on the mass transfer and Laplace pressure framework, $\mathrm{{f_{\infty \chi}},{f_{0\chi}}} [-]$ are the asymptotical and percolation threshold invaded fractions respectively, $\mathrm{\overline{\tau}_{\chi}[s]}$, $\mathrm{\Delta \overline{p}_\chi (t)}\mathrm{[N/m^2]}$ are the mean mass transfer timescale and mean Laplace pressure as a function of time, $\phi$ is the cumulative distributive function (CDF) of Gaussian distribution.  The corresponding fitting parameters are $\mathrm{\mu_{t\chi}}$,$\mathrm{\sigma_{t\chi}}$ and $\mathrm{\mu_{p\chi}}$,$\mathrm{\sigma_{p\chi}}$. 
The subscript $\mathrm{`\chi'}$ denotes the specific realization.

For a specific interface, $\mathrm{I}$, we define the mass transfer timescale, $\mathrm{\tau_{I}}$, as the ratio of the characteristic length scale, $\mathrm{L_{char}(=(\frac{V}{S})_{I})}$ to mass transfer coefficient, $\mathrm{k_{I}}$ at the onset of secondary invasion.

\begin{equation}
\mathrm{\tau_{I}=\frac{(\frac{V}{S})_{I}}{k_{I}}    }
\label{eq:19}
\end{equation}

We then define the mean mass transfer timescale of $\mathrm{\chi^{th}}$ realization associated with the interfaces at the onset of secondary invasion, as follows:

\begin{equation}
\mathrm{\overline{\tau}_{\chi}} = \left( \frac{1}{\mathrm{J}} \sum_{\mathrm{I}=1}^{\mathrm{J}} \left( \mathrm{\frac{\left( \frac{V}{S} \right)_{I}}{k_{I}} }\right) \right)_{\chi}
\label{eq:20}
\end{equation}

We define the Laplace pressure, $\mathrm{\Delta p}$ as the difference between the local hydrodynamic pressure and the local capillary pressure. 
The mean at each timestep is defined as follows:

\begin{equation}
\mathrm{\Delta \overline{p}_\chi (t)} = \left( \mathrm{\overline{P}(t) - \overline{p}(t)} \right)_{\chi}
\label{eq:21}
\end{equation}

Each ensemble is generated using a shape parameter of the Gamma distribution, consists of at least 500 network configurations. 
The skewness factor of each realization, $\mathrm{Sk}$, is defined as the median/mean ratio of the pore throats. 
At the ensemble level, we define a fitting efficiency metric, which depicts the number of realizations having $\mathrm{R^2 > 0.9}$, based on the choice of MT or LP approach (eq. \eqref{eq:17} and \eqref{eq:18}, respectively). 
The representative fit of each ensemble is the realization with the highest $\mathrm{R^2}$ value. The quantities that are normalized are scaled relative to their respective values at the lowest skewness factor (\( \hat{a} = \frac{\overline{a}}{\overline{a}_{E1}} \)).

\section{Results and Discussion}

To explore the difference between mass transfer and Laplace pressure based analysis, we study the spatio-temporal evolution of fluid flow in an anomalous pore network. 
The anomalous nature is measured in terms of skewness, generated using a Gamma distribution function that accounts for the mean size ($\mathrm{\overline{r}}$) and the shape parameter. 
The higher the value of the shape parameter, the more symmetric the distribution becomes. 
We generate 1000 random network configurations, produced from the same distribution function with a pre-determined mean pore throat size and shape parameter; thus, each configuration will have a slightly different mean pore throat size and variance, while the ensemble of all the configurations follows the PDF's prescribed mean and skewness, as summarized in table~\ref{table:3}. 

 We consider the replacement of the water invaded into the oil-saturated system by surfactant-enriched water, which leads to the secondary invasion. To begin the simulations, we choose the inlet pressure to be the lowest capillary pressure of the pore throats, resulting in an initial invaded fraction, referred to as the percolation threshold. For the sake of brevity, we will refer to this stage of invasion, before the surfactant effect sets in, as the primary invasion. 
Similarly, we refer to the secondary invasion as the stage where the surfactant is introduced and begins to accumulate at the interfaces.

\begin{figure*}
%\begin{sidewaysfigure*}
\centering
\includegraphics[width=0.8\linewidth]{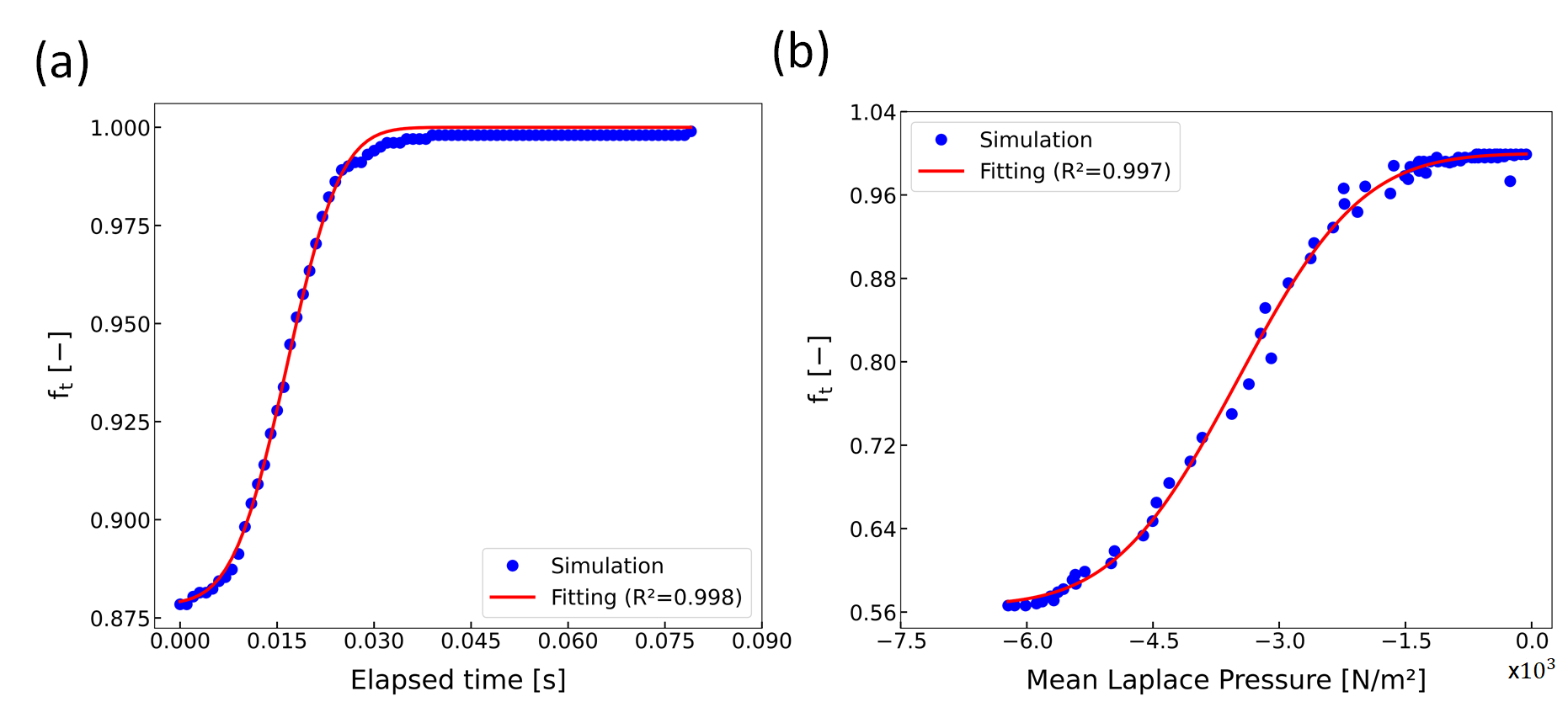}
\caption{Invaded fraction, $\mathrm{f_t}$, and accompanying fits as a function of (a) elapsed time and (b) mean Laplace pressure. These results are obtained for the most anomalous distribution (ensemble E1), which has the lowest fitting efficiencies for the ensemble due to their high skewness (83.7$\%$ and 81.4$\%$ for MT and LP analysis, respectively). 
The corresponding mean pore throat sizes are $50.2,51.1[\mu\text{m}]$, with variance of $1288.9, 1264.9 [\mu\text{m}^2]$. The invaded fraction continues to evolve due to surfactant adsorption to the oil-water interface (bulk concentration, $\mathrm{C_{B}}=2\,\text{mM}$ for surfactant S1), depicting important stages of invasion dynamics.}
\label{fig:3} 
%\end{sidewaysfigure*}
\end{figure*}

In Fig.~\ref{fig:3}, we show the spatio-temporal evolution of the pore invasion by calculating the invaded fraction as a function of time and Laplace pressure. 
We pick a bulk concentration to start the simulation, based on the initial invasion pattern established by a prescribed percolation threshold, and select the parameter values of the surfactant, $\mathrm{S1}$ (Table~\ref{table:3} in Section~\ref{section:Supplementary}). 
As can be seen from the plot, the secondary invasion is triggered right away, as the invasion fraction rises above the initial percolation threshold value. 
Note that the percolation threshold values are different for both plots because they are representative fits obtained after employing MT and LP based approaches. 

\begin{figure}[H]
%\begin{sidewaysfigure*}
\centering
\includegraphics[width=0.8\linewidth]{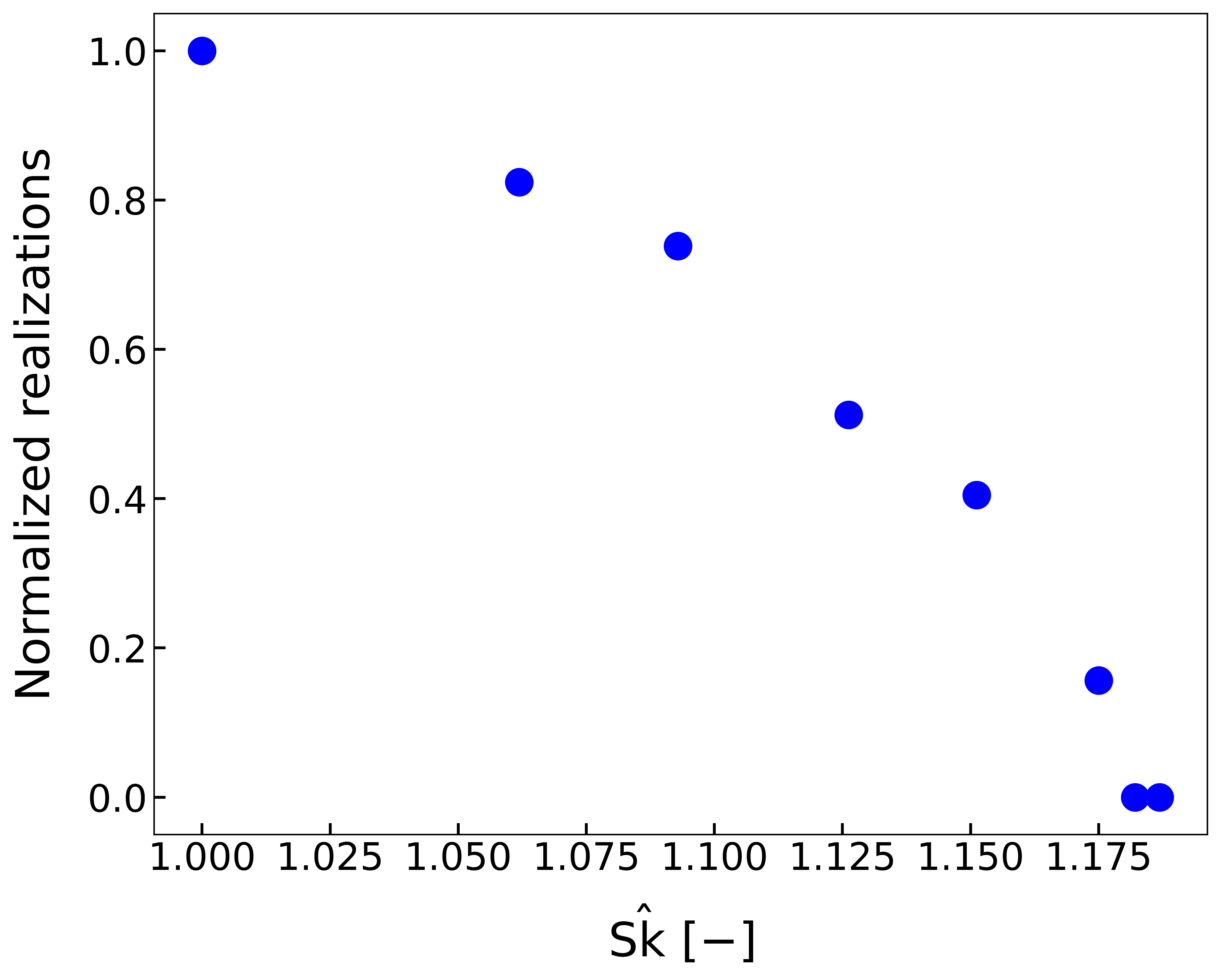}
\caption{Normalized realizations of each ensemble where the defending fluid was not fully displaced i.e. incomplete displacement as a function of normalized mean skewness factor, $\mathrm{\hat{Sk}}$.}
\label{fig:16} 
%\end{sidewaysfigure*}
\end{figure}

The next stage is marked by the continuous accumulation of more surfactants at the nearby interfaces and at the interfaces newly formed by the invasion. 
Every time a new pore throat is invaded, due to the changed balance between the IFT and the local Capillary pressure induced by the surfactant, the mass transfer coefficient, k, of the remaining active interfaces, is updated. 
This process continues until the system reaches a steady state, which occurs when the entire network is invaded by water, and oil is fully displaced from the system, or when additional changes in IFT are no longer sufficient to invoke additional invasions. 
However, for some ensembles, there were realizations where the oil was not fully displaced (Fig.~\ref{fig:16}, in Section~\ref{section:Supplementary}). 
In order to maintain a consistent comparison among the $\mathrm{Sk}$ values, and as the $\mathrm{Sk}$ decreases, we only focus on fully invaded realizations, where trapped ganglia are reduced through the network boundary or out-of-plane nodes (see Fig. \ref{fig:1}), while realizations with trapped ganglia will be considered in a future study. 

\begin{figure}[H]
%\begin{sidewaysfigure*}
\centering
\includegraphics[width=0.8\linewidth]{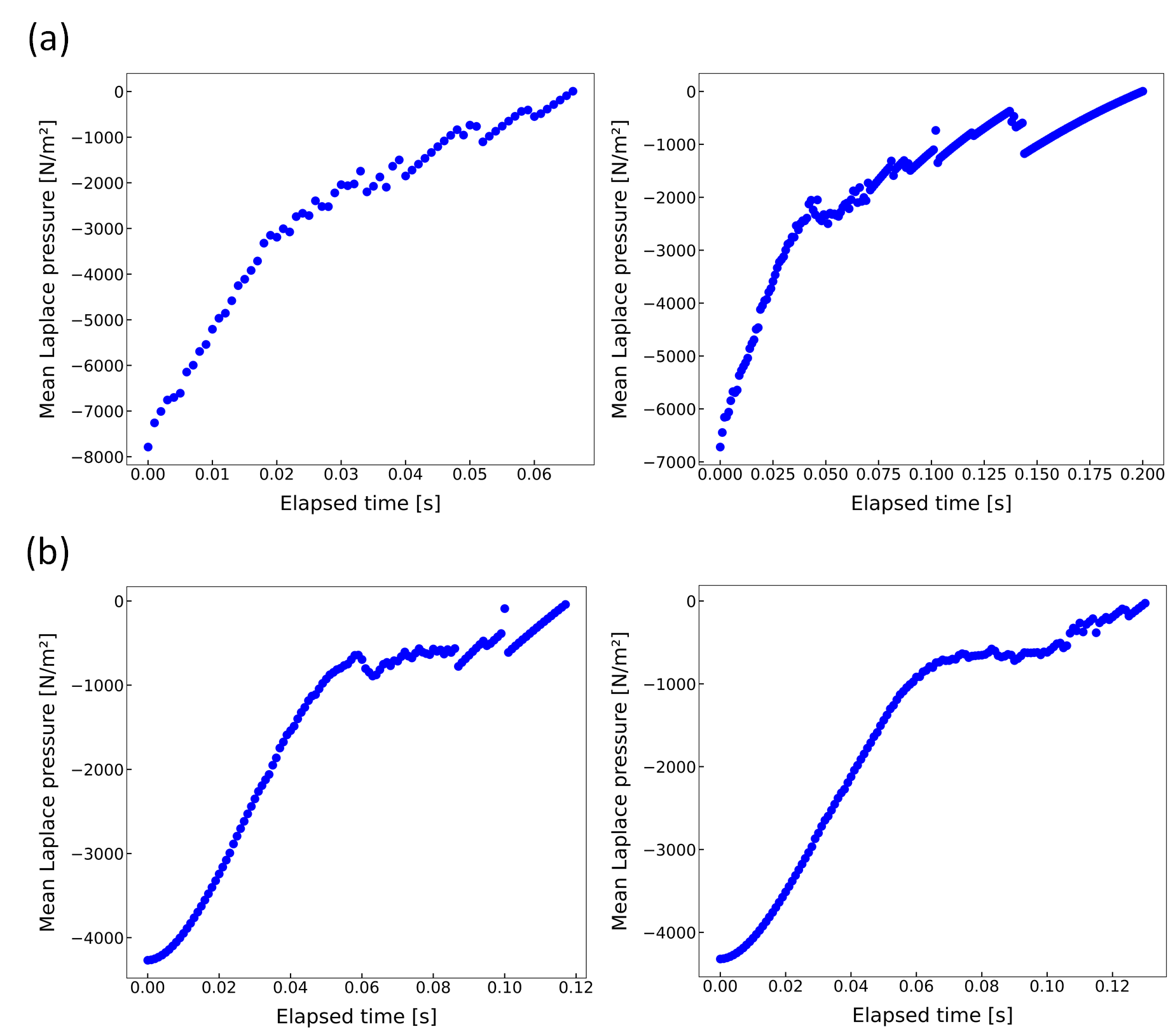}
\caption{Mean Laplace pressure with respect to time of two random realizations in ensembles (a) E1 and (b) E8, depicting a heavy tail. In E1, the transition is relatively steeper compared to that in E8 (note the change in elapsed time range between E1 and E8).   }
\label{fig:17} 
%\end{sidewaysfigure*}
\end{figure}

\begin{figure}[H]
%\begin{sidewaysfigure*}
\centering
\includegraphics[width=0.8\linewidth]{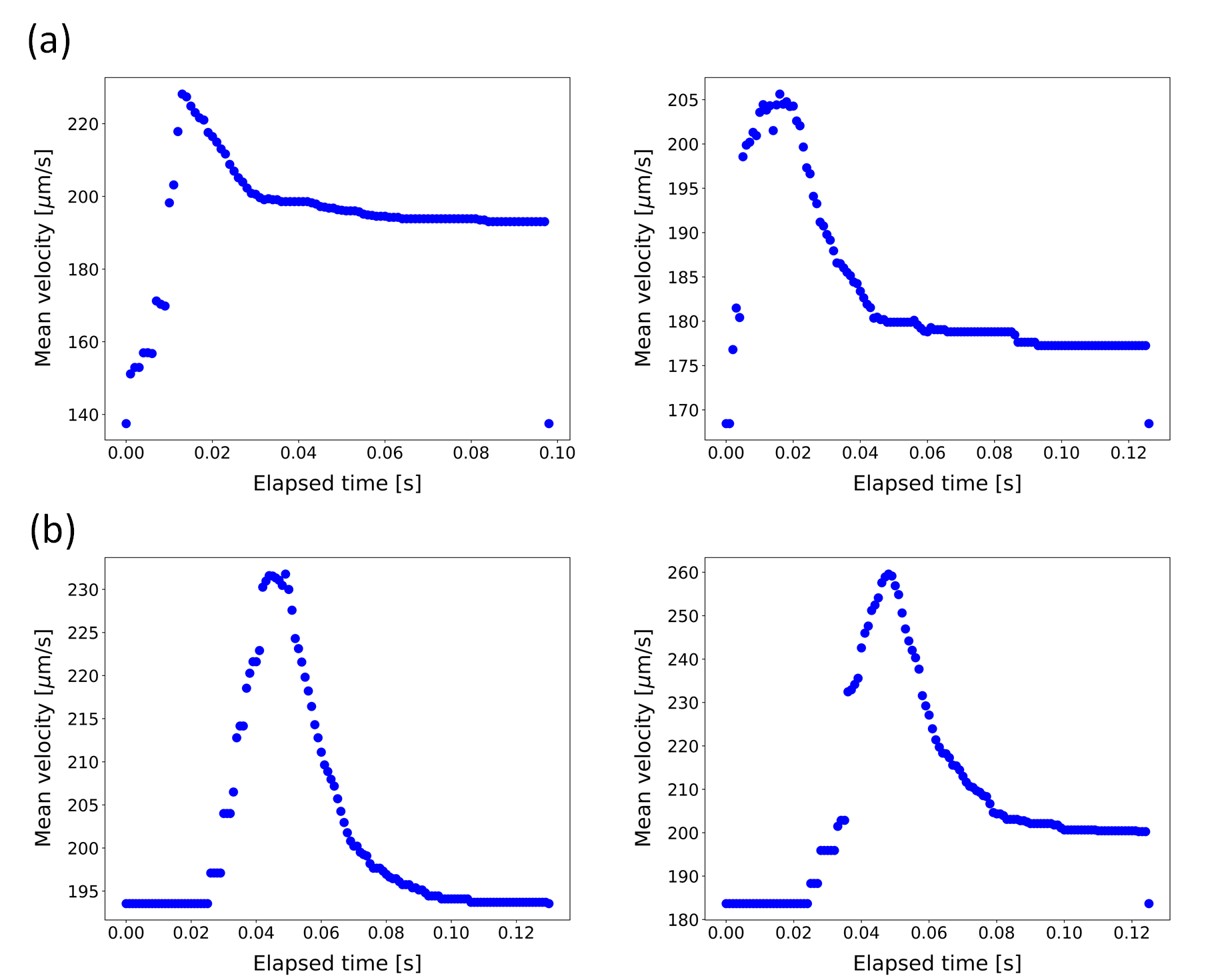}
\caption{Mean velocity with respect to time of two random realizations in ensembles (a) E1 and (b) E8, displaying a mix of Gaussian-like behavior, either anomalous or symmetric, respectively.}
\label{fig:18} 
%\end{sidewaysfigure*}
\end{figure}

In our PNM simulations, the pore throats follow a Gamma distribution, spatially. 
This pore radii distribution establishes not only the capillary pressure for each node, but also the surfactant adsorption by establishing the velocity distribution for the invaded phase via the Hagen-Poiseuille equation (eq.\eqref{eq:4}). 
As such, the increased invaded fraction over time can be considered as the cumulative update of the velocity field change. 
For the prescribed conditions in our simulation, namely fixed pressure difference, the additional invasions reduce the overall resistance of the PNM, and therefore increase the overall flux. Following this change in the velocity field and overall flux within the PNM, we use the CDF of the Gaussian PDF, to juxtapose the spatio-temporal invasion dynamics of each realization using the mass transfer based (eq.~\ref{eq:17}) and Laplace pressure based approach (eq.~\ref{eq:18}). 
Furthermore, we introduce the mass transfer timescale at the secondary stage, defined as a ratio of the characteristic length scale to mass transfer coefficient (eq.~\ref{eq:19}). 
The characteristic length scale is taken to be the inverse of the surface area-to-volume ratio. 
Dividing this length scale by the mass transfer coefficient provides a quantitative measure of the rate at which mass transfer occurs from the bulk to the interface. 
Combining this time scale with the Gaussian CDF, namely the Error function, allows us to juxtapose the invasion fraction evolution in time with a known set of parameters. 
On similar grounds, we calculate the mean Laplace pressure at each timestep and divide it by the inlet pressure. 
This provides insight into the characteristic behavior of the system from another perspective. 
We also observe the anomalous behavior of Laplace pressure as a function of time for all ensembles (Fig.~\ref{fig:17} in Section~\ref{section:Supplementary}). 
However, the velocity distribution follows a Gaussian-like behavior across different ensembles (Fig.~\ref{fig:18} in Section~\ref{section:Supplementary}).  

Since each realization has a different percolation threshold value but the same steady state value, calculating all the relevant parameters at the end of primary invasion/onset of secondary invasion stage ensures that the analysis takes into account the initial condition. As shown in Fig.~\ref{fig:3}, the fitting equations (eq.~\ref{eq:17},~\ref{eq:18}) behave reasonably well for the representative realization of the MT and LP approaches for ensemble E1. 
It is important to note here that since each network configuration is independent yet produced using an identical distribution, it is not possible to solve the integral in the error function analytically. Instead, one has to solve each realization numerically.  

We repeat the described fitting for the surfactant-driven invasion dynamics for each realization with a bulk concentration and parameters of surfactant S1. 
Each realization has different percolation threshold values, confirming that the pore throat sizes are randomly arranged. 
Another noteworthy point is that regardless of the percolation threshold, the realizations always lead to an invaded fraction value of 1.0 at steady state. 
This means that the initial condition does not influence the maximum invaded fraction. 

\begin{figure*}
    %\begin{sidewaysfigure*}  % Uncomment this line if you want the figure rotated sideways
    \centering
    \includegraphics[width=0.8\linewidth]{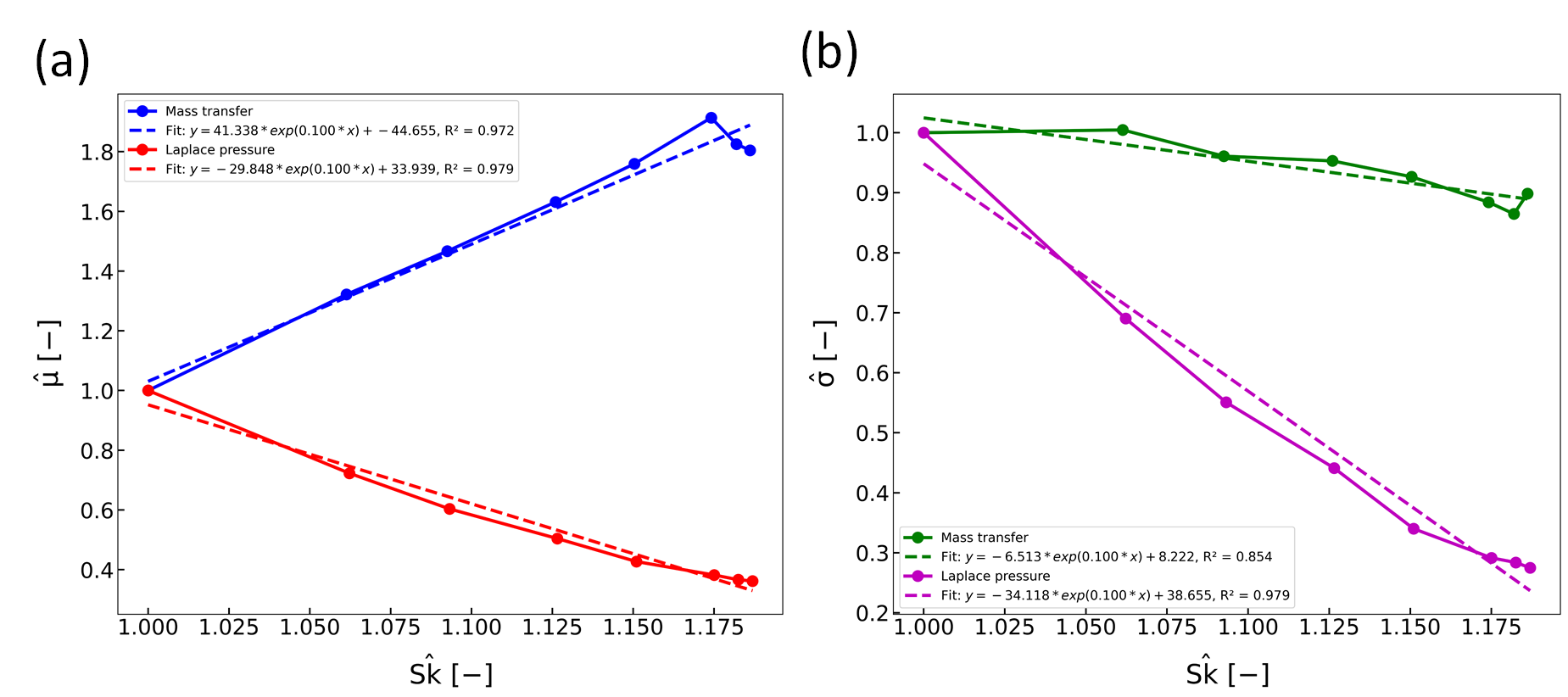}
    \caption{
    Normalized fitting parameters, $\mathrm{\hat{\mu}}$ and $\mathrm{\hat{\sigma}}$ obtained from the fitting are plotted against normalized skewness, $\mathrm{\hat{Sk}}$, comparing the Mass transfer based and Laplace pressure based approaches for a large number of realizations with surfactant, S1. Mass transfer tends to become increasingly dominant relative to the Laplace pressure, owing to the rising trend of $\mathrm{\hat{\mu}}$ and the comparatively slower decline of $\mathrm{\hat{\sigma}}$ relative to the corresponding trend in Laplace pressure.
    }
    \label{fig:4} 
    %\end{sidewaysfigure*}  % Uncomment this line if you want the figure rotated sideways
\end{figure*}

\begin{figure}[H]
%\begin{sidewaysfigure*}
\centering
\includegraphics[width=0.8\linewidth]{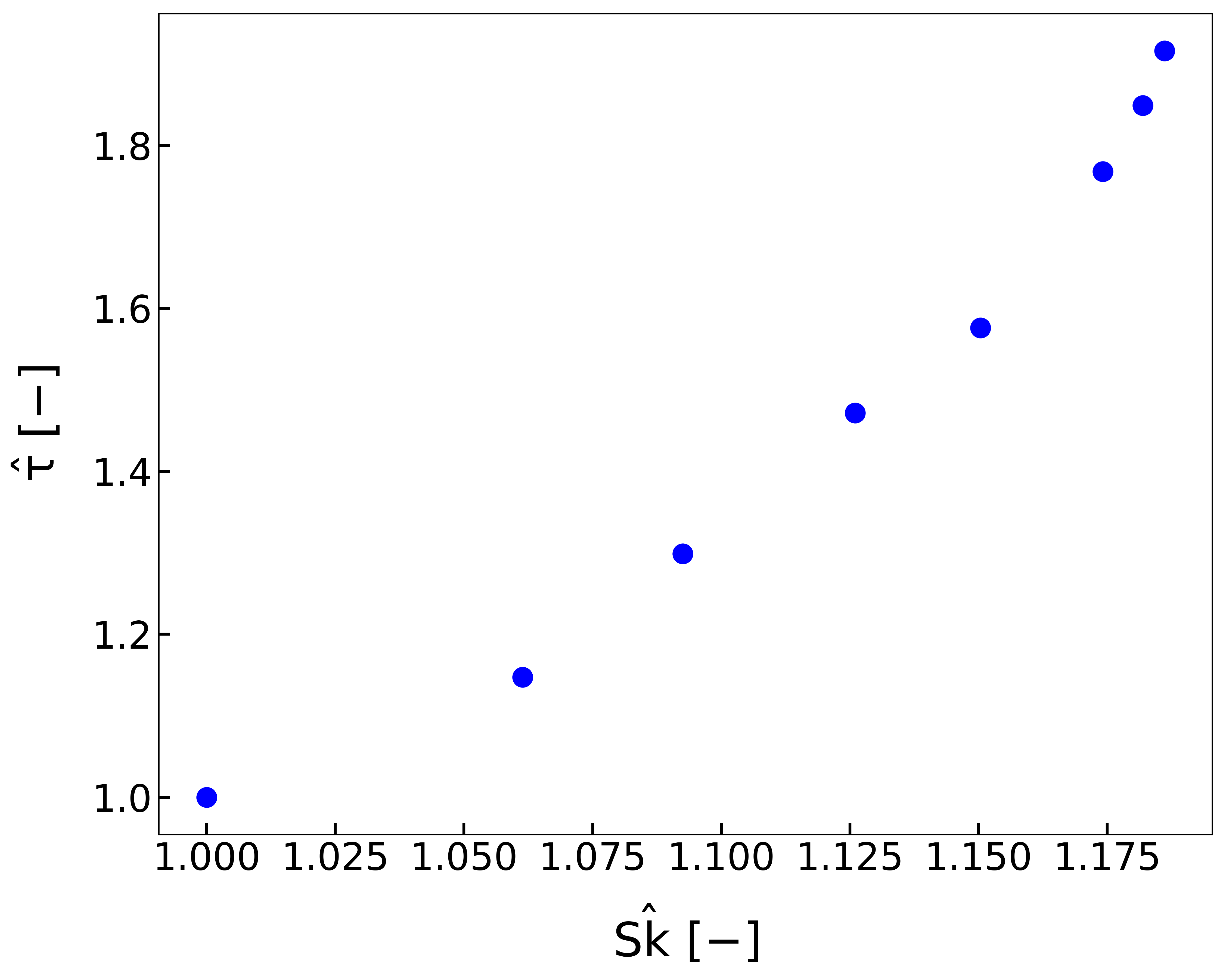}
\caption{Normalized mean mass transfer timescale, $\hat{{\tau}}$ at the onset of secondary invasion at the ensemble level, $\mathrm{\overline{\tau}}$ as a function of normalized mean skewness factor, $\mathrm{\hat {Sk}}$.}
\label{fig:11} 
%\end{sidewaysfigure*}
\end{figure}

\begin{figure}[H]
%\begin{sidewaysfigure*}
\centering
\includegraphics[width=0.8\linewidth]{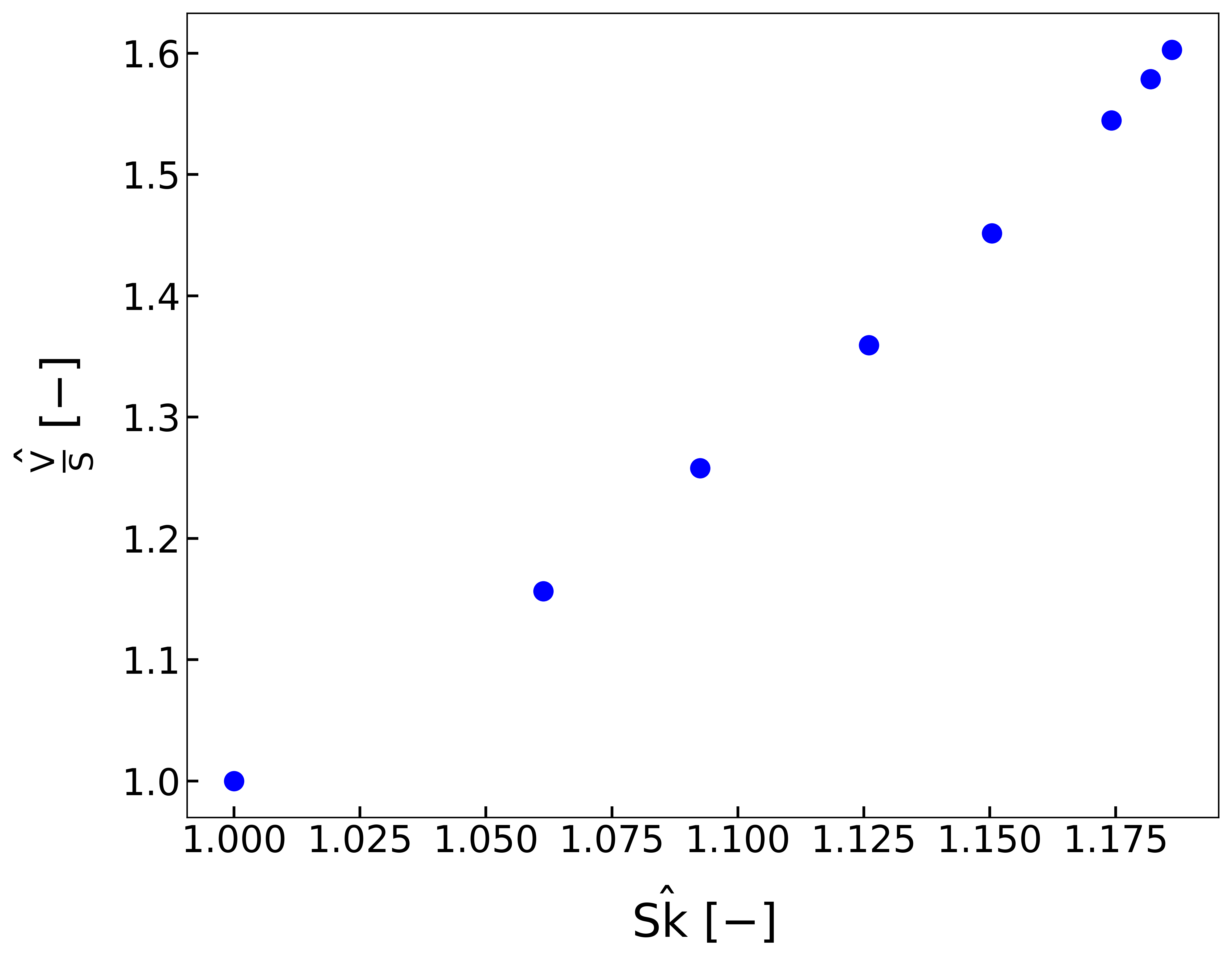}
\caption{Normalized mean Volume-to-Surface area ratio, $\mathrm{\hat{\frac{V}{S}}}$ at the onset of secondary invasion at the ensemble level,  as a function of normalized mean skewness factor, $\mathrm{\hat {Sk}}$.}
\label{fig:12} 
%\end{sidewaysfigure*}
\end{figure}

\begin{figure}[H]
%\begin{sidewaysfigure*}
\centering
\includegraphics[width=0.8\linewidth]{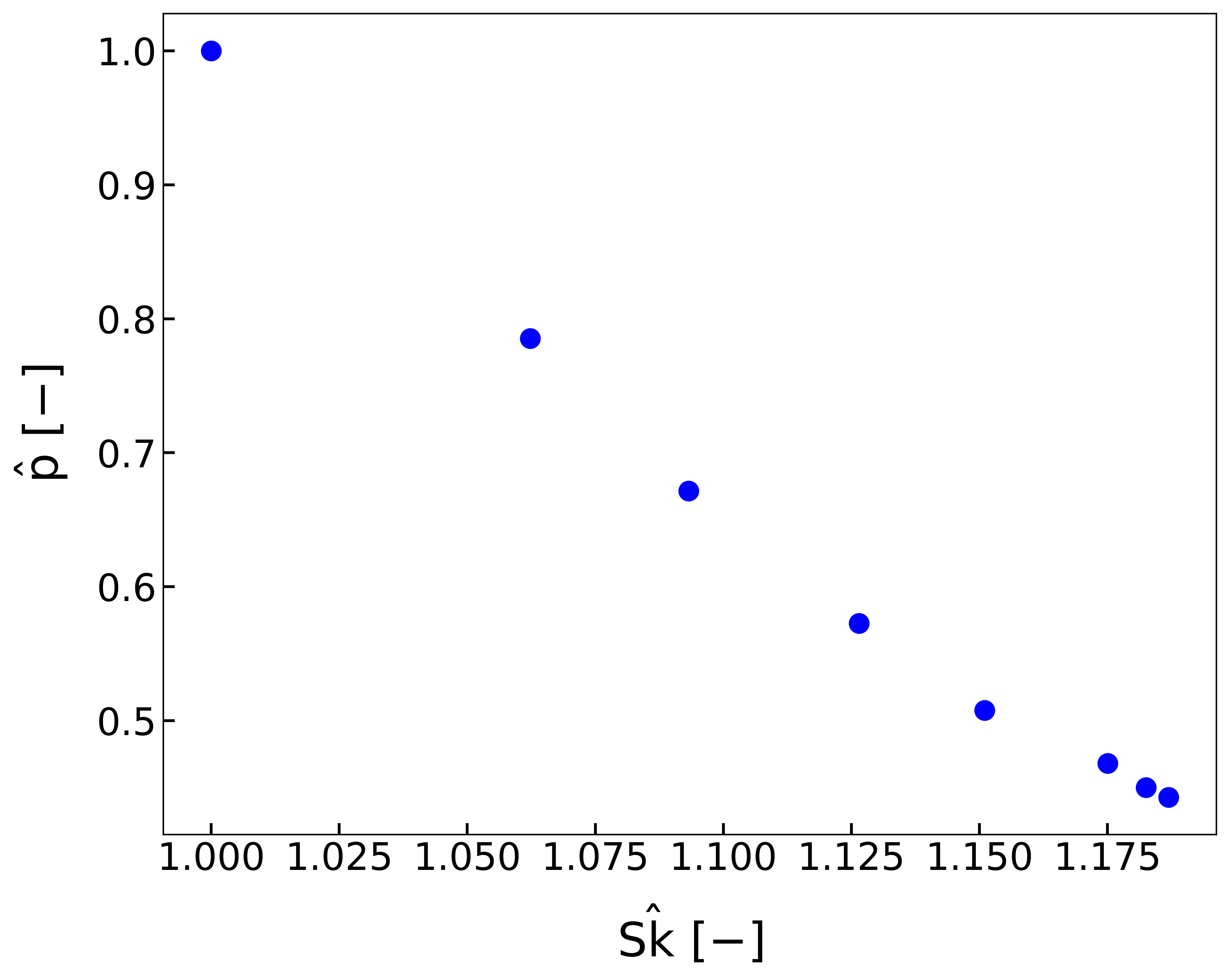}
\caption{Normalized mean capillary pressure, $\mathrm{\hat p}$  at the onset of secondary invasion at the ensemble level as a function of normalized mean skewness factor, $\mathrm{\hat{Sk}}$.}
\label{fig:13} 
%\end{sidewaysfigure*}
\end{figure}

\begin{figure}[H]
%\begin{sidewaysfigure*}
\centering
\includegraphics[width=0.8\linewidth]{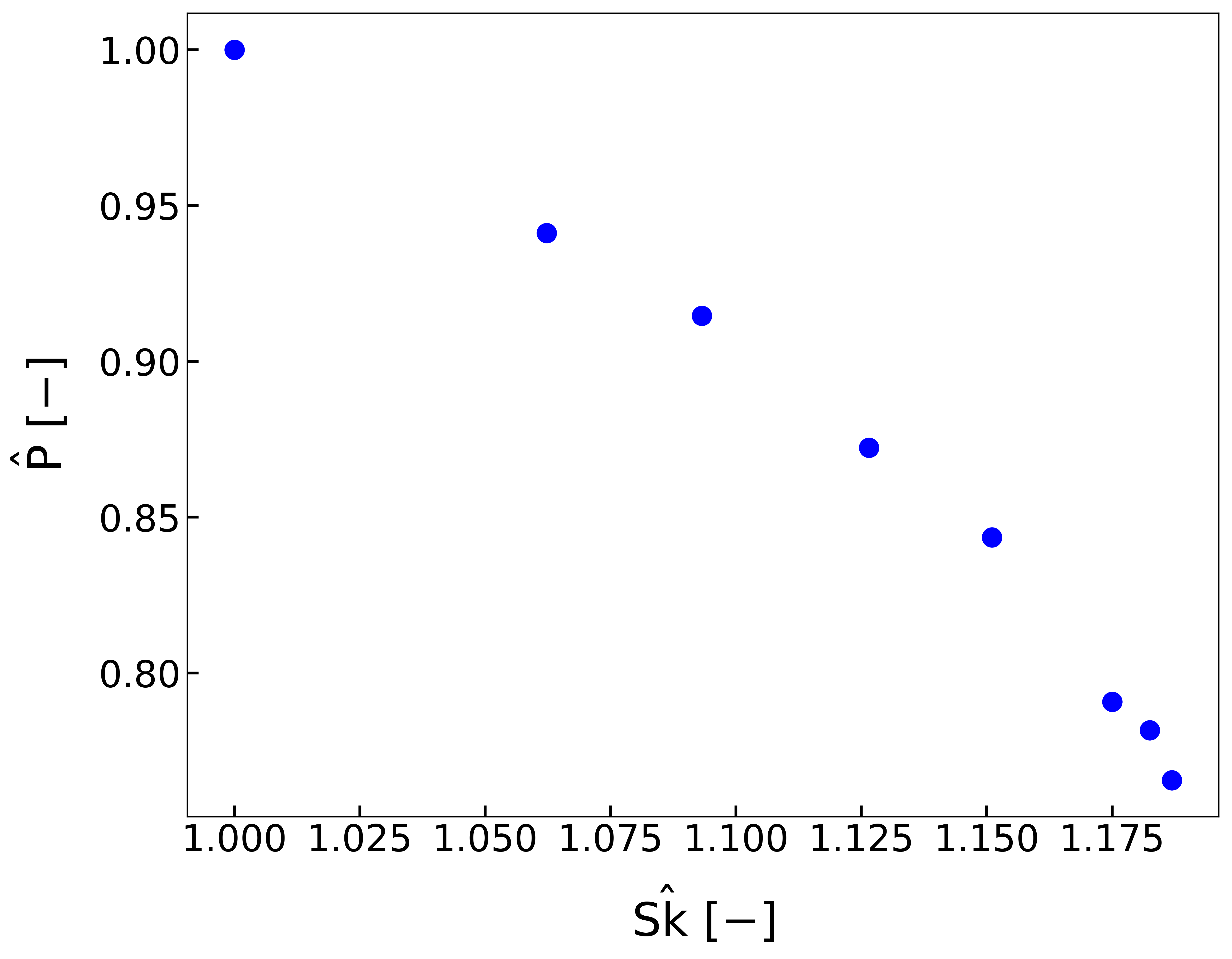}
\caption{Normalized mean hydrodynamic pressure, $\mathrm{\hat P}$ at the onset of secondary invasion at the ensemble level,  as a function of normalized mean skewness factor, $\mathrm{\hat {Sk}}$.}
\label{fig:14} 
%\end{sidewaysfigure*}
\end{figure}

From Fig.~\ref{fig:4}(a), to enable a direct comparison between the two approaches, we employ an exponential fitting function with a fixed rate constant. As can be seen, their behaviors differ: the MT fitting parameter increases while in the LP case, it decreases with increasing skewness. 
This can be attributed to the fact that the mass transfer timescale increases with skewness, owing to the higher Volume-to-Surface area ratio (Fig.~\ref{fig:11},~\ref{fig:12} in Section~\ref{section:Supplementary}).  
The decreasing trend in the LP case can be explained by looking at the capillary pressure and hydrodynamic pressure. 
The former decreases more sharply compared to the latter with increasing skewness, as is evident from the plots (Fig.~\ref{fig:13},~\ref{fig:14} in Section~\ref{section:Supplementary}). 

As the skewness increases, the pores tend to become more uniform in size. This leads to a decrease in the variability of the Laplace pressure which is shown in Fig.~\ref{fig:4}(b). 
Symmetry reduces the extreme variations in pore sizes, which means the pore network becomes more uniform. 
Laplace pressure, which is sensitive to curvature (and thus pore size variation), is impacted more strongly as the network becomes more symmetric. The reduced skewness towards larger pores, as the network becomes symmetrical, forms higher capillary pressures which require stronger reduction in IFT for the secondary invasion. 
The reduced variability of mass transfer, compared with Laplace pressure, comes from the uniform behavior of the flow paths, resulting in a larger timescale.

\section{Conclusion}

In this study, the surfactant-driven spatio-temporal invasion dynamics in anomalous pore networks are investigated by integrating the commonly used pore-network model with a simplified surfactant transport and adsorption model.  
Model calculations are focused on the evolution of fluid invasion due to the alteration of IFT by surfactant adsorption. 
We employ two complementary approaches to analyze the emerging dynamics: Mass transfer based analysis (MT), where we introduce a mass transfer timescale, and Laplace pressure based analysis (LP), where we use the inlet pressure to scale the changing Laplace pressure. 
Both approaches allow for a detailed understanding of the invasion process from different perspectives, highlighting the role of surfactant adsorption in shaping the overall dynamics.

%Our results reveal a significant trend: as the symmetry of the pore throat distribution increases, the influence of Laplace pressure begins to diminish, and mass transfer starts to dominate the dynamics. This shift suggests that in more symmetric systems (less anomalous), where the pore throat sizes converge, the mass transfer characteristics become more uniform and predictable, leading to a more consistent invasion process. On the other hand, in highly anomalous systems, Laplace pressure and capillary effects play a dominant role in determining the flow behavior, as the larger variations in pore size create a broader range of flow velocities. 

Our results reveal the following trend: as the system becomes more anomalous, the Laplace pressure at the onset of secondary invasion tends to become more positive, and a broader range of flow velocities decreases the mass transfer time scale, therefore prolonging the secondary invasion time, $\mathrm{\hat{\mu}}$, mirroring the capillary effects marked by the mean Laplace pressure. 
In contrast, in less anomalous systems — where both pore throat size variations and volume-to-surface are larger — mass transfer dominates the dynamics. 
The mass transfer reflects the uniform sorption coverage of the surfactant, leading to nonlinear progression toward the critical Laplace pressure required for invasion. 
This trend highlights how the increasing anomalous nature of the pore throat size distribution and volume-to-surface ratio, which leads to non-uniform surfactant sorption, enhances the role of capillary forces by broadening the target Laplace pressure and mass transfer time. 
Many rock formations and soils exhibit large heterogeneity ranges, with non-symmetrical pore size distributions; our analysis is crucial for understanding how these distributions of pore sizes can influence the overall dynamics of surfactant-driven invasion, as it highlights the correlation between the mass transfer and Laplace pressure-based approaches.   

\section{Appendix}
\label{section:Supplementary}

\subsection{Adjacency matrix and accuracy of the PNM}

\begin{figure}[H]
\centering
\includegraphics[width=0.8\linewidth]{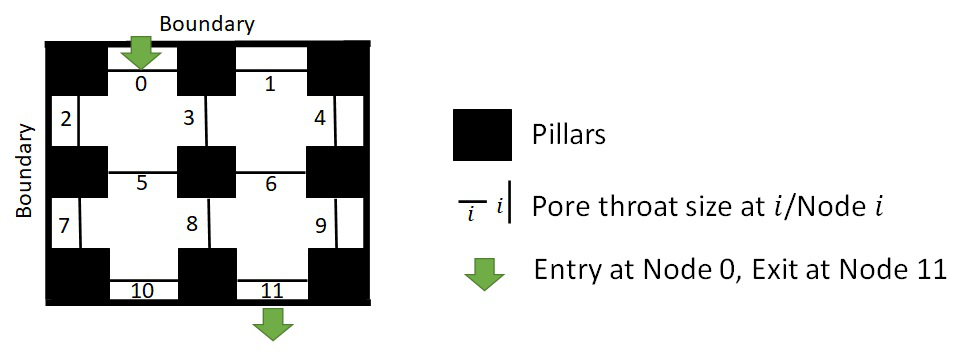}
\caption{A $3 \times 3$ pillar array depicting the inlet and outlet nodes with the numbering scheme.}
\label{fig:5} 
\end{figure}

\begin{figure}[H]
\centering
\includegraphics[width=0.8\linewidth]{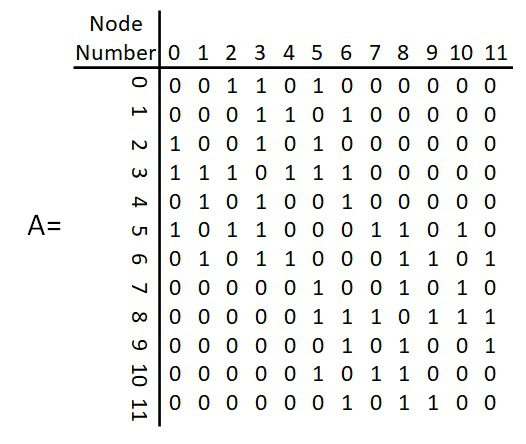}
\caption{Adjacency matrix, $\mathrm{A}$, of nodes of the $3 \times 3$ pillar array.}
\label{fig:6} 
\end{figure}

For an $\mathrm{N} \times \mathrm{N}$ pillar array, the number of pore throats is $2\mathrm{N(N-1)}$ and the number of pore bodies is $\mathrm{(N-1)^2}$. We consider a $3 \times 3$ pillar array which has 12 pore throats and 4 pore bodies (Fig.~\ref{fig:5}). The adjacency matrix of this pillar array is shown in Fig.~\ref{fig:6}. Clearly, each node has 3 or 6 connections. In a similar fashion, we generate adjacency matrix of much larger pillar arrays, ensuring that each node has 3 or 6 connections.

\begin{figure}[H]
\centering
\includegraphics[width=0.8\linewidth]{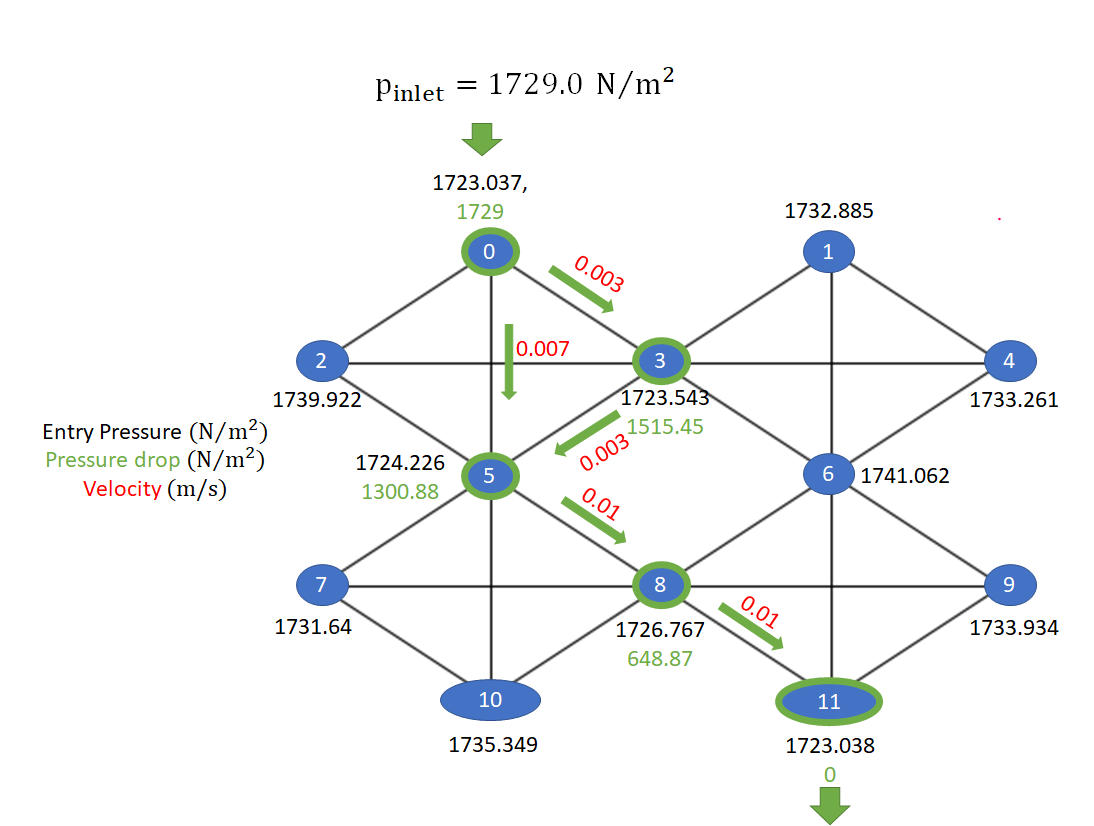}
\caption{Connectivity of the $3 \times 3$ pillar array along with the viscous pressure drop values and velocity directions for \(\text{p}_{\text{inlet}} = 1729.0 \, \text{N/m}^2\). The nodes highlighted in green indicate that they are invaded.}
\label{fig:7}
\end{figure}

\begin{figure}[H]
\centering
\includegraphics[width=0.8\linewidth]{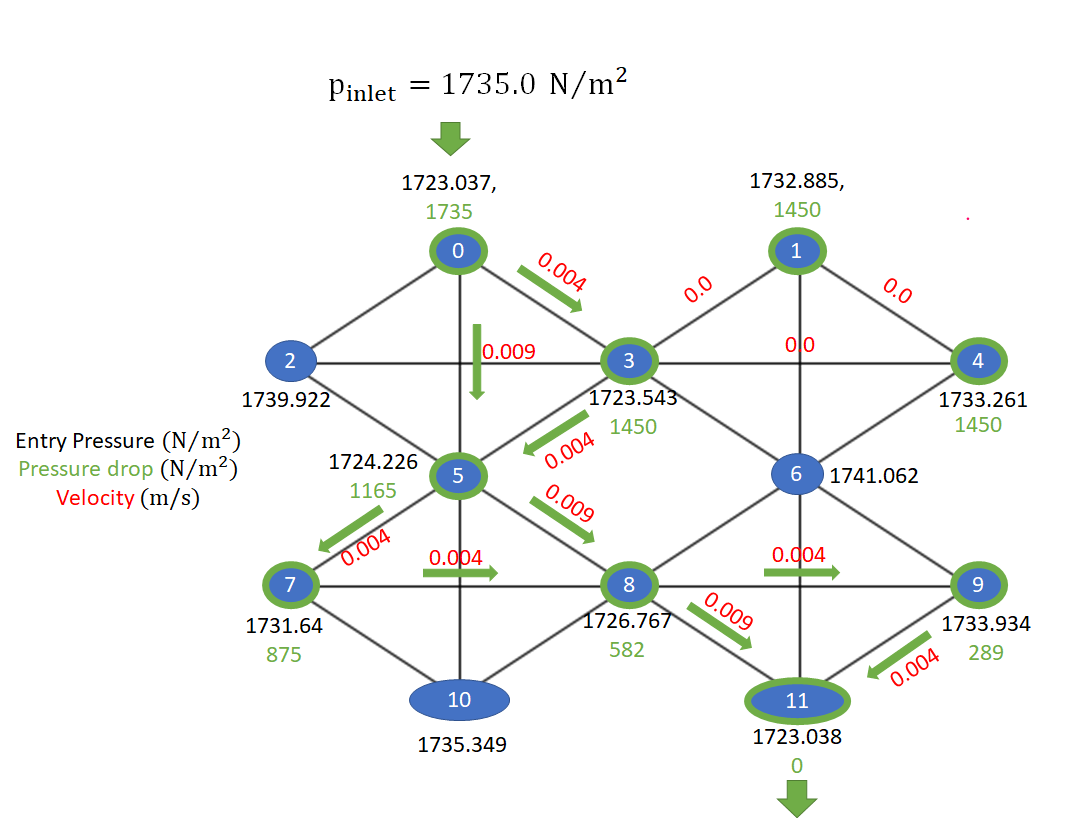}
\caption{Connectivity of the $3$x$3$ pillar array along with the viscous pressure drop and velocity directions for $\mathrm{p_{inlet}}=1735.0$ $\mathrm{N/m^2}$. The nodes highlighted in green means that they are invaded.}
\label{fig:8} 
\end{figure}

To check the accuracy of the invasion percolation algorithm, we choose an inlet pressure $\mathrm{p_{inlet}} =1729.0$ $\mathrm{N/m^2}$. From Fig.~\ref{fig:7}, we see which nodes are invaded based on the criterion that the inlet pressure must be greater than the local capillary pressure for invasion to happen. Since the invasion pattern is established now, we calculate the viscous pressure drop and the velocities of each connected node. We increase the inlet pressure to $\mathrm{p_{inlet}} =1735.0$ $\mathrm{N/m^2}$ and we see more nodes are invaded and calculate the viscous pressure drop and velocities of the connected nodes Fig.~\ref{fig:8}.

\subsection{Surfactant Parameters}

\begin{table}[H]
\centering
\includegraphics[width=0.8\linewidth]{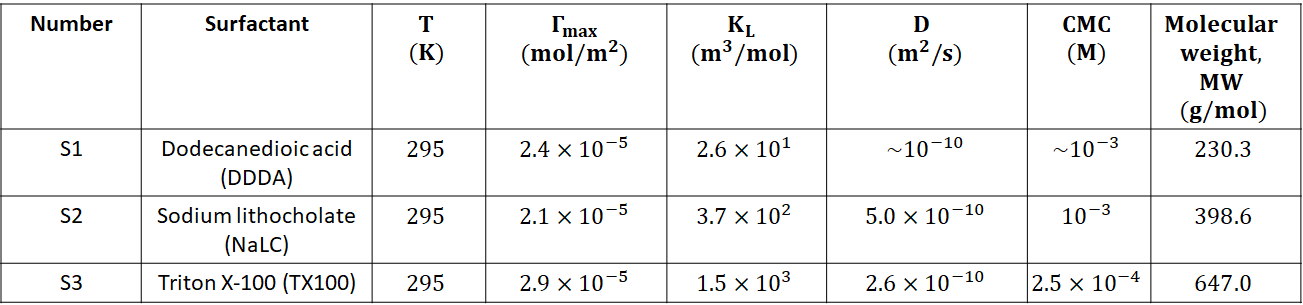}
\caption{Langmuir adsorption equilibrium parameters of various surfactants in aqueous solution \cite{Chang1995}.}
\label{table:3} 
\end{table}

\subsection{Effect of surfactant concentration on invasion dynamics}

\begin{figure}[H]
%\begin{sidewaysfigure*}
\centering
\includegraphics[width=0.8\linewidth]{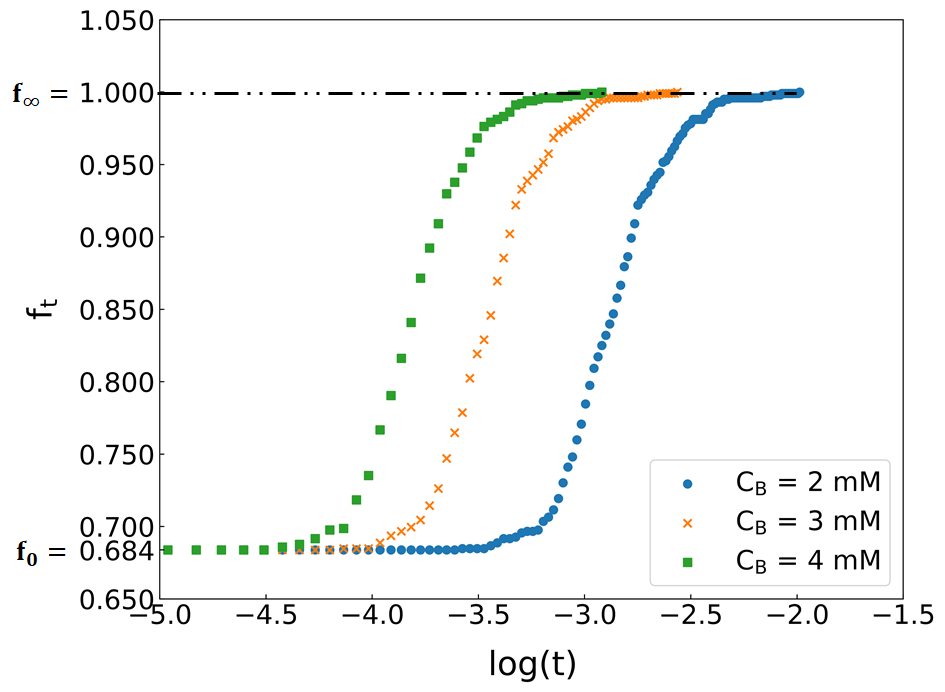}
\caption{Invaded fraction as a function of time for $\mathrm{C_{B}}=2,3,4\,\text{mM}$, $\mathrm{\overline{r}}=49.895\,\mu\text{m}$, and $\mathrm{var}=4.813 \,\mu\text{m}^2$.}
\label{fig:9} 
%\end{sidewaysfigure*}
\end{figure}

When $\mathrm{C_{B}}$ is increased, we find that the secondary invasion triggers earlier and the dynamics reach the steady state much faster. The results are shown in Fig.~\ref{fig:9}. This happens because as the bulk concentration is higher, the local bulk concentration to every interface is higher. This causes more surfactants to be adsorbed at the interface in a relatively short time. As a result, the IFT changes more quickly and the invasion proceeds further. 

\subsection{Effect of surfactant properties on invasion dynamics}

Since we are considering Langmuir adsorption isotherm, the model employs the following surfactant parameters: Temperature, $\mathrm{T}$, maximum adsorption capacity, $\mathrm{\Gamma_{max}}$, Langmuir equilibrium constant, $\mathrm{K_{L}}$, diffusion coefficient, $\mathrm{D}$ and molecular weight, $\mathrm{MW}$, borrowed from literature for different surfactants \cite{Chang1995}. Since surfactant migration from the bulk to the interface is a complex process where both diffusion and advection play a role, we choose a simplified quantity called mass transfer coeffcient, $\mathrm{k}$. This leads to a local concentration close to every interface which adsorbs at the interface. The degree to which this quantity will adsorb depends on the maximum adsorption capacity of the interface ($\Gamma_{max}$) and the binding affinity governed by Langmuir equilibrium constant, $\mathrm{K_{L}}$. Our goal in this paper is to show that the combined PNM-SAM is applicable for a wide range of binding affinities, $\mathrm{K_{L}}$. To achieve this, we pick different surfactants $\mathrm{S1, S2, S3}$, all of which have $\mathrm{K_{L}}$ over a wide range. 

\begin{figure}[H]
%\begin{sidewaysfigure*}
\centering
\includegraphics[width=0.8\linewidth]{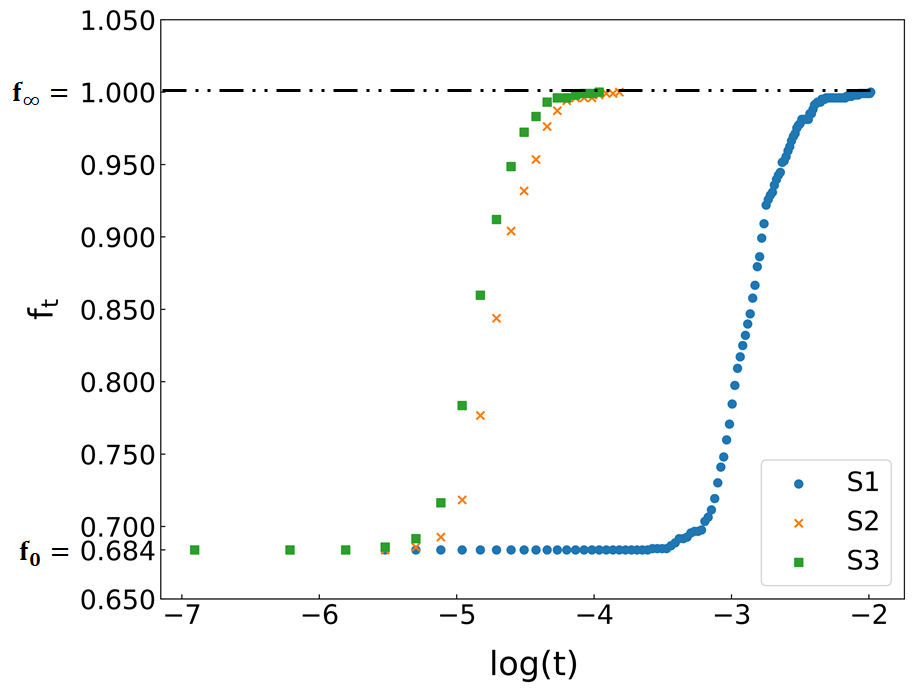}
\caption{Invaded fraction as a function of time for $\mathrm{C_{B}}=2\,\text{mM}$, $\mathrm{\overline{r}}=49.895\,\mu\text{m}$, and $\mathrm{var}=4.813 \,\mu\text{m}^2$ for three different surfactants $\mathrm{S1,S2,S3}$. The corresponding surfactant parameters are mentioned in Table~\ref{table:2}.}
\label{fig:10} 
%\end{sidewaysfigure*}
\end{figure}

As can be seen from Fig.~\ref{fig:10}, the invaded fraction versus time for surfactant, $\mathrm{S3}$ is more expedited compared to $\mathrm{S1}$. We are keeping $\mathrm{C_{B}}$ and the network configuration the same in each case. The expedited nature of $\mathrm{S3}$ can be attributed to not only higher $\mathrm{K_{L}}$ but other model parameters as well which are surfactant dependent. Thus, it is relatively difficult to isolate a single model parameter which causes this behavior. 

\subsection{Mass transfer (MT) based plots}

\subsection{Laplace Pressure (LP) based plots}

\begin{figure}[H]
%\begin{sidewaysfigure*}
\centering
\includegraphics[width=0.8\linewidth]{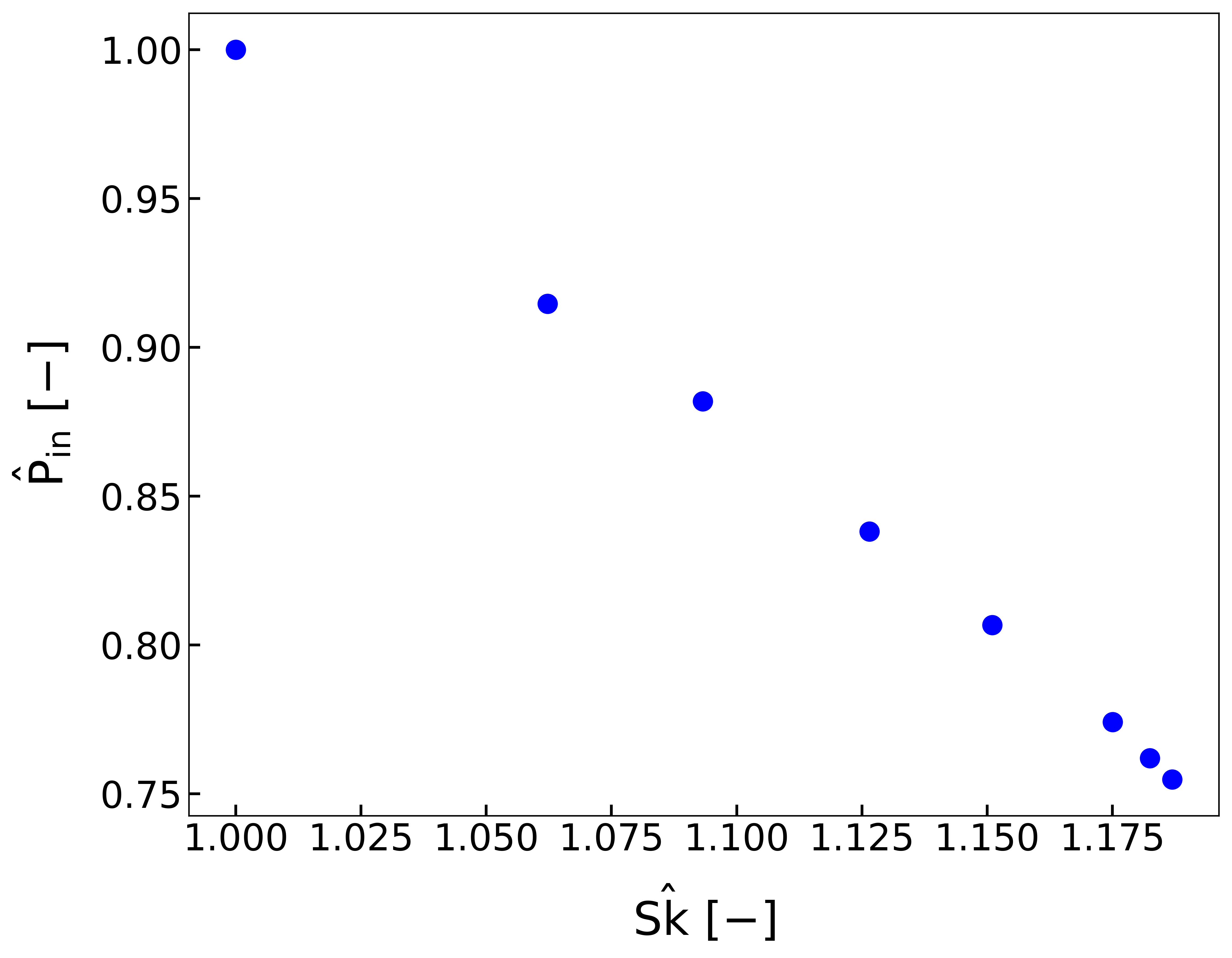}
\caption{Normalized mean inlet pressure at the ensemble level, ${\mathrm{{\hat P_{in}}}}$, as a function of normalized mean skewness factor, $\mathrm{\hat{Sk}}$.}
\label{fig:15} 
%\end{sidewaysfigure*}
\end{figure}

\subsection{Incomplete displacement}

\subsection{Anomalous nature of Laplace pressure}

\subsection{Gaussian-like nature of velocity distribution}

\bibliography{Ref}

\end{document}